\documentclass{iopart}
\usepackage{graphicx}
\usepackage[light]{draftcopy}

\begin{document}
\title{Vetoes for Inspiral Triggers in LIGO Data}


\author{Nelson Christensen$^1$\footnote{nchriste@carleton.edu},
Peter Shawhan$^2$\footnote{shawhan\_p@ligo.caltech.edu}
and
Gabriela Gonz\'alez$^3$\footnote{gonzalez@lsu.edu},
For the LIGO Scientific Collaboration}
\address{$^1$Physics and Astronomy, Carleton College,
Northfield, MN 55057, USA\\
$^2$LIGO Laboratory, California Institute of Technology, Pasadena, CA 91125 USA \\
$^3$Department of Physics and Astronomy, Louisiana State University,
Baton Rouge, LA 70803, USA\\
}

\date{\today}


\begin{abstract}
Presented is a summary of studies by the LIGO Scientific 
Collaboration's Inspiral Analysis Group on the development of possible
vetoes to be used in evaluation of data from the first two LIGO
science data runs.
Numerous environmental monitor signals and interferometer control 
channels have been analyzed in order to characterize the interferometers' 
performance. The results of studies on selected data segments are 
provided in this paper. The vetoes used in the compact binary inspiral
analyses of LIGO's S1 and S2 science data runs are presented and
discussed.
\end{abstract}



\section{Introduction} 
\label{intro}

The Laser Interferometer Gravitational Wave Observatory (LIGO) is now operating, and
collecting meaningful scientific data~\cite{LIGO-DET}. The LIGO Scientific Collaboration (LSC)
is conducting searches for several types of gravitational-wave signals.
To date, analysis of data from LIGO's first science data run has led to
the publication of searches for
continuous waves from pulsars~\cite{LIGO-CW}, the ``inspiral'' (orbital decay)
of compact binary systems~\cite{LIGO-IN}, short bursts~\cite{LIGO-BU}, 
and an isotropic stochastic background~\cite{LIGO-ST}.  

The waveform emitted by an inspiraling compact binary system can be
modeled accurately (at least if the component masses are fairly low),
allowing the use of matched filtering techniques when searching for
this class of signals.  The data is filtered using a large number of
``template'' waveforms in order to search for signals with a
range of physical parameters.  For any given template,
the search algorithm generates a ``trigger''
each time the output of the matched filter exceeds a pre-determined
threshold in signal-to-noise ratio (SNR), provided that the frequency
distribution of the signal power is consistent with the expected
waveform, checked quantitatively using a $\chi^2$ test.

\begin{figure}[tb]
  \begin{center}
    \includegraphics[width=5.0in,angle=0]{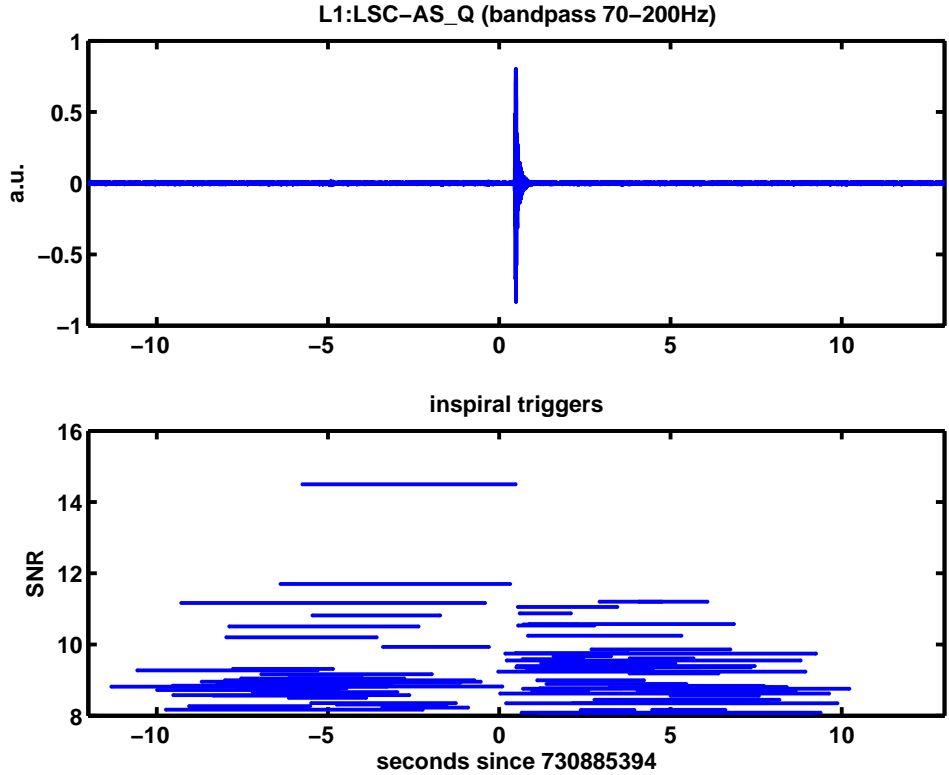}
  \end{center}
\caption{An example of how a large amplitude {\it glitch} can cause
numerous templates to report significant SNR triggers. The top trace
shows a glitch observed in the time series of the LIGO Livingston
gravitational wave channel, denoted L1:LSC-AS\_Q.  The bottom plot
shows the inspiral triggers with SNR$> 8$ which were reported (based
on filtering with many template waveforms) in the vicinity of this
glitch.  Each trigger is represented by a horizontal bar which extends
from the time at which the template waveform passes 100 Hz to the
inferred coalescence time.  The vertical position of the bar indicates
the maximum SNR observed in that template.  The inferred coalescence
times extend over a span of $\sim$$16$~s.} \label{fig1}
\end{figure}

\begin{figure}[tb]
  \begin{center}
    \includegraphics[width=5.0in,angle=0]{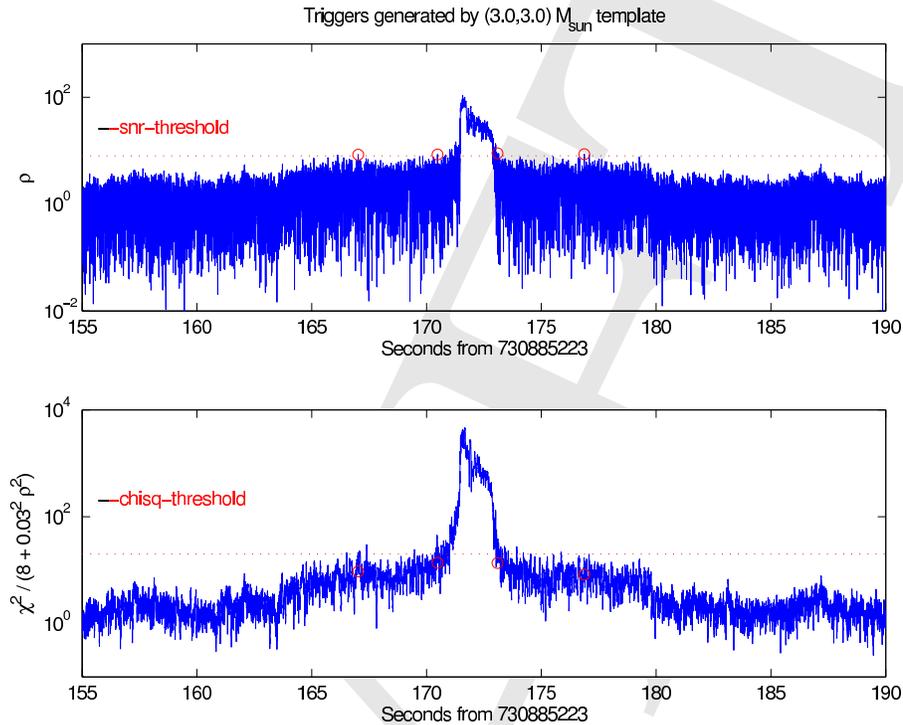}
  \end{center}
\caption{Time series displays of the output of the matched filter, for
one particular template, in the vicinity of the large glitch shown in
Figure~\ref{fig1}.  The top plot shows the SNR and the threshold used to
identify triggers, while the bottom plot shows the $\chi^2$ variable
which is required to be below a threshold.
The circles note the times when the signal exceeded the SNR threshold (top),
yet passed the $\chi^2$ test (bottom).
} \label{fig2}
\end{figure}

While this
search algorithm is optimal in the case of stationary Gaussian noise,
the actual noise in the LIGO interferometers is strongly influenced by
optical alignment, servo control settings, and environmental
conditions.
Large amplitude {\it glitch} events, or short stretches of increased
broadband noise, will excite the inspiral filter for many templates, thereby
leading to false triggers in the search. An example of this can be
seen in Figure~\ref{fig1}, where a large-amplitude glitch causes
numerous inspiral templates to respond over a time span as long as
$\sim$$16$~s.  This time scale is related to the treatment of sharp
features in the power spectral density of the detector noise, which is
used as an inverse weighting factor in the matched filter.
Figure~\ref{fig2} shows the output of the matched filter in the
vicinity of this glitch, illustrating how these
inaccurate inspiral coalescence times can 
arise from the ringing of the template filter:
although the main SNR peak is easily rejected by the $\chi^2$ test,
there are a few nearby times for which the SNR exceeds the trigger
threshold while $\chi^2$ is below the rejection threshold.

The goal of the studies described in this paper is to eliminate
demonstrably bad stretches of data and to identify environmental or
instrumental causes of glitches when possible, allowing us to ``veto''
(reject) any inspiral triggers occurring at nearby times.  In addition
to the main data channel in which a gravitational wave signal would
appear (called ``LSC-AS\_Q'' because it is the Length Sensing and
Control signal extracted from the ``Anti-Symmetric port'' photodiode
using Quadrature demodulation phase), numerous additional channels are
recorded to monitor auxiliary optical signals and servo control points
in the interferometer, as well as environmental conditions.  In some
cases, we are able to significantly reduce the rate of false triggers
by using these additional channels as indicators of instrumental or
environmental disturbances.

LIGO's first science data run, called S1, spanned 17 days from August
23 to September 9, 2002.  The second science data run, called S2,
spanned two months from February 14 to April 14, 2003.  The average
noise in the LIGO interferometers was roughly an order of magnitude
better during S2 than during S1.  Building on the analysis of the S1
data~\cite{LIGO-IN}, a
search for binary neutron star (BNS) inspiral events is being conducted with the S2 data; 
an upper limit will be placed on the coalescence rate in the Milky Way
and nearby
galaxies~\cite{S2}.
The specifics determining the vetoes are presented in the remainder of the paper. 
Section~\ref{S1} outlines the concepts used veto studies and
summarizes the S1 veto analysis; a more complete description can
be found in \cite{LIGO-IN}. A comprehensive description of the S2 inspiral veto analysis is presented 
in Section~\ref{S2}. A summary of our conclusion, and thoughts on possible future analysis 
plans, is contained in Section~\ref{disc}. In the course of this paper we refer to the 
4 km interferometer 
at Livingston, Louisiana, as L1, and the 4 km and 2 km interferometers at Hanford, 
Washington, as H1 and
H2 respectively. 

\section{Vetoes for LIGO Science Data Run S1}
\label{S1}

A description of the vetoes implemented for the BNS inspiral 
analysis of data from LIGO science data run
S1 \cite{LIGO-IN} is presented here.
In order to avoid the possibility of statistical bias, potential veto
conditions were studied using only a ``playground'' data set 
comprising about 10~\% of the collected data, selected by hand to give
a sampling of different degrees of non-stationarity observed in the
detector noise at different times.
This playground data 
was not used in the calculation of the inspiral rate limit.

Only the L1 and H1 interferometers were used for the S1 inspiral
analysis.  For either interferometer, sections of data were excluded
from examination if there were problems with calibration signals.
This resulted in 5~\% of L1 data being excluded, and 7~\% of the H1 data.
In addition, periods of time when the noise level of an interferometer
was abnormally large were excluded from analysis. This determination 
was made through the monitoring
of the band-limited root-mean-square noise that 
occurred in four frequency bands \cite{LIGO-IN,LIGO-BU}.
This veto eliminated 8~\% of L1 data and 18~\% of H1 data.

Numerous interferometer control and environmental monitoring channels were examined at times when the
inspiral templates reported triggers during the playground section,
in order to look for correlations.  The subset of channels which
showed a possible correlation were processed using a glitch-finding
program which generated ``veto triggers''.  These veto triggers were
compared to the list of inspiral triggers, with an adjustable time
window to account for instrumental delays as well as the different
trigger generation algorithms.
The effectiveness of a channel as a veto, using a given time window,
was measured by calculating the veto efficiency (fraction of inspiral
triggers rejected by veto triggers), usage fraction (fraction of veto
triggers coincident with at least one inspiral trigger), and dead-time
(fraction of total run time during which inspiral triggers would be
rejected according to the set of veto triggers and the time window).

The H1 channel H1:LSC-REFL\_I, a photodiode signal at the
interferometer's reflected port, was found to contain large glitches
which correlated well with large glitches seen in the
gravitational-wave channel.  A program called {\it
glitchMon}\footnote{{\it glitchMon}, written by M.\ Ito (University
of Oregon), is a program which looks for transient signals in selected
LIGO data channels.  It is based on the LIGO Data Monitoring Tool
(DMT) library.} was used to filter the H1:LSC-REFL\_I channel and
record large excursions as veto triggers.  A time window of $\pm 1$~s
around these veto trigger times yielded a veto efficiency of over
60~\% for inspiral triggers with SNR~$>10$, with
a deadtime of only 0.2~\%.  A prospective veto condition for the L1
interferometer, using a channel called L1:LSC-AS\_I which is derived
from the same photodiode as the gravitational-wave channel, was
abandoned due to concerns that a gravitational wave could appear in
this channel with non-negligible amplitude.

Once these data quality and veto conditions had been developed using
the playground data, they
were subsequently implemented as part of the S1 analysis
pipeline~\cite{LIGO-IN}. Inspiral triggers that
passed the SNR threshold, $\chi^2$ test, and veto condition were
reported as event candidates and were used to calculate an upper limit
on the rate of binary inspirals in the Galaxy.  A ``{\it
  post-mortem}'' examination of these events
provided illuminating information. For example, the ``loudest'' event
detected in the L1 data
was the result of a saturation of the interferometer's antisymmetric port photodiode, probably caused 
by a misalignment in the optical system. These results, and the experience from the S1 veto analysis
served as a starting point for the examination of the S2 data.

\section{Vetoes for LIGO Science Data Run S2}
\label{S2}

The character of the S2 data was very different from that of S1. 
The stability of all of the LIGO interferometers had improved significantly, and the 
quality of the data was dramatically improved.
The interferometer sensitivities had also
improved, and consequently new noise sources became visible. The experience derived from the 
S1 analysis was brought forward, but due to the different behavior of the interferometers
it was necessary to reinspect all of the interferometer control and environmental monitoring
channels in detail again. Numerous tools were used for the task. What was initially helpful was to 
use the inspiral template triggers, found in playground data, 
and to inspect candidate channels at these times.

Data quality examinations (more comprehensive than those done for the
S1 analysis) provided the means to exclude sections of data where there were obvious
problems. A number of problems caused data to be excluded:
data outside of the official S2 run times, missing data, missing or unreliable calibration,
non-standard servo control settings (in a few L1 segments), and input/output controller timing problems at L1.
The playground data was then used to judge the relevance of other
potential data quality flags, leading to two additional data quality vetoes.
One concerned the H1 interferometer, which suffered from occasional
episodes of elevated non-stationary broadband noise.  We eliminated
data in which the noise level in the upper part of the sensitive
frequency band was high for consecutive periods of at least three
minutes; this requirement ensured that a real gravitational wave
inspiral signal would not invoke this veto condition, even if it had
an exceptionally large amplitude.
The other data quality veto used pertained to the saturation of the photodiode at the 
antisymmetric port at any of
the LIGO interferometers, as was observed during S1. 
This effect correlated with a small, but significant number of L1 inspiral 
triggers.

As in the S1 veto study, numerous channels, with various filters and
thresholds, were processed with {\it glitchMon} to produce veto triggers.
The efficiency and dead-time for each possible veto condition was
evaluated using a playground data set, which for the S2 run consisted
of 600 seconds out of every 6370 seconds of data.  This definition of
the playground ensured that it was representative of the entire run;
for instance, it included some data from all times of the day.
The ``safety'' of several potential veto channels was evaluated by
injecting simulated gravitational-wave signals into the interferometer
arm lengths and checking for the signals to appear in various
auxiliary channels.  The signals were found to appear in just one
tested channel, L1:LSC-AS\_I, with measurable amplitude, so that
channel was deemed to be unsafe for use as a veto.

No good candidate veto channels were identified for H1 and H2,
however, there were a few candidates for L1.
Non-stationary noise in low frequency part of the sensitivity range
used for inspiral search, initially $50$--$2048$~Hz,
appeared to be dominant cause for deleterious glitches in the data.
In particular, the non-stationary noise in L1 had dominant frequency
content around $70$~Hz.
A key auxiliary channel,  L1:LSC-POB\_I , also had highly variable noise at $70$~Hz. 
There are understandable physical mechanisms for this:  the power recycling servo loop 
(for which L1:LSC-POB\_I is the error signal) has a known 
instability around 70~Hz when the gain is too high; independently,
when the gain of the differential arm length servo loop goes too low (due to low optical gain), 
glitches around 70~Hz tend to appear. Sometimes
these glitches in L1:LSC-POB\_I couple into the differential arm length
signal sufficiently strongly to produce inspiral triggers.
To avoid these excess triggers, we decided to increase the
lower bound of the frequency band used for the BNS inspiral search to 100~Hz. 
This reduced the number of inspiral triggers, and simulations
indicated the loss of sensitivity for the target population of binary
neutron star systems was acceptably small.

\begin{figure}[tb]
  \begin{center}
    \includegraphics[width=5.0in,angle=0]{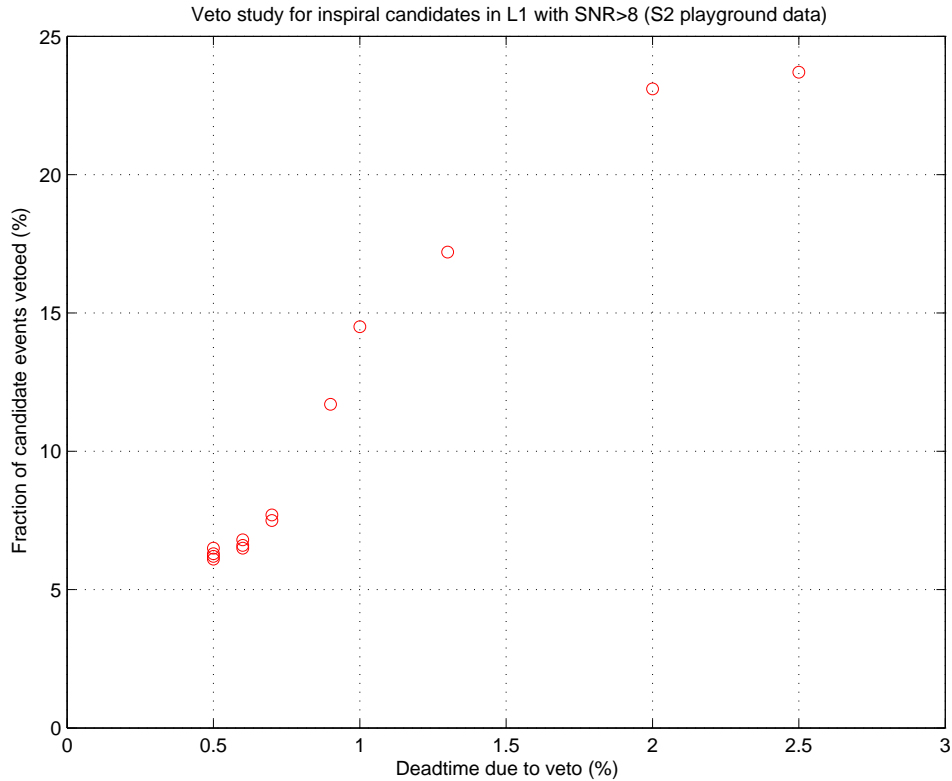}
  \end{center}
\caption{An example of the 
veto efficiency (for BNS inspiral triggers in L1)
versus deadtime for the veto channel L1:LSC-POB\_I, 
using symmetric windows of $0.0, \pm 0.05, \pm 0.1, \pm 0.15, 
\pm 0.2, \pm 0.25, \pm 0.3, \pm 0.4, \pm 0.5, \pm 0.75, \pm 1.0, \pm 2.0, \pm 4.0$, 
and $\pm6.0$~s.
The data from L1:LSC-POB\_I was filtered with a fourth order Chebyshev 70~Hz high-pass filter,
and excursions found by {\it glitchMon} with significance of $7
\sigma$ or greater were taken to be veto triggers.
These results are from the S2 playground data.
} \label{fig3}
\end{figure}

The lists of veto triggers produced by {\it glitchMon} were compared
to the output of the inspiral template search using 
data in the S2 playground.
Figure~\ref{fig3} shows an example of the 
veto efficiency versus dead-time for the channel L1:LSC-POB\_I, using
symmetric time windows of $0.0, \pm 0.05, \pm 0.1, \pm 0.15, 
\pm 0.2, \pm 0.25, \pm 0.3, \pm 0.4, \pm 0.5, \pm 0.75, \pm 1.0, \pm 2.0, \pm4.0$ and
$\pm6.0$~s.  In this case, the L1:LSC-POB\_I data was filtered with a
fourth-order Chebyshev high-pass filter with a corner frequency of
70~Hz, and {\it glitchMon} triggers with a significance of more than 
$7 \sigma$ were taken to be veto triggers.
Note that the veto efficiency rises significantly as the time window
is increased.
As was illustrated in Figure~\ref{fig1}, a large-amplitude glitch can
cause the inspiral search algorithm to generate triggers with inferred
coalescence times rather far from the time of glitch.  For this
reason, we found that we had to use rather long veto time windows to
achieve good veto efficiency.  After a long series of studies, we
settled on using the L1:LSC-POB\_I channel, with the filtering and
threshold given above, with a very wide and asymmetric window,
$-4$~s to $+8$~s.
In the playground data, this veto condition vetoed 27~\% of the BNS
inspiral triggers with SNR~$>8$, and 35~\% of the inspiral triggers
with SNR~$>10$, with a dead-time of 2.5~\%. The usage fraction of the
veto was 25~\% for SNR~$>8$ and 7~\% for SNR~$>10$, while the expected
random use would be 4.6~\% and 0.5~\% respectively.  The final
analysis of the full S2 data set (excluding the playground) was done
using a more stringent $\chi^2$ threshold to reduce the number of
false triggers, so the final veto efficiencies and usage fractions are
somewhat lower than the numbers given above: the efficiency is
13~\% for inspiral triggers with SNR~$>8$ and $30 \pm 10$~\% for
inspiral triggers with SNR~$>10$, with dead-time of 3.0~\%.

\begin{figure}[tb]
  \begin{center}
    \includegraphics[width=5.0in,angle=0]{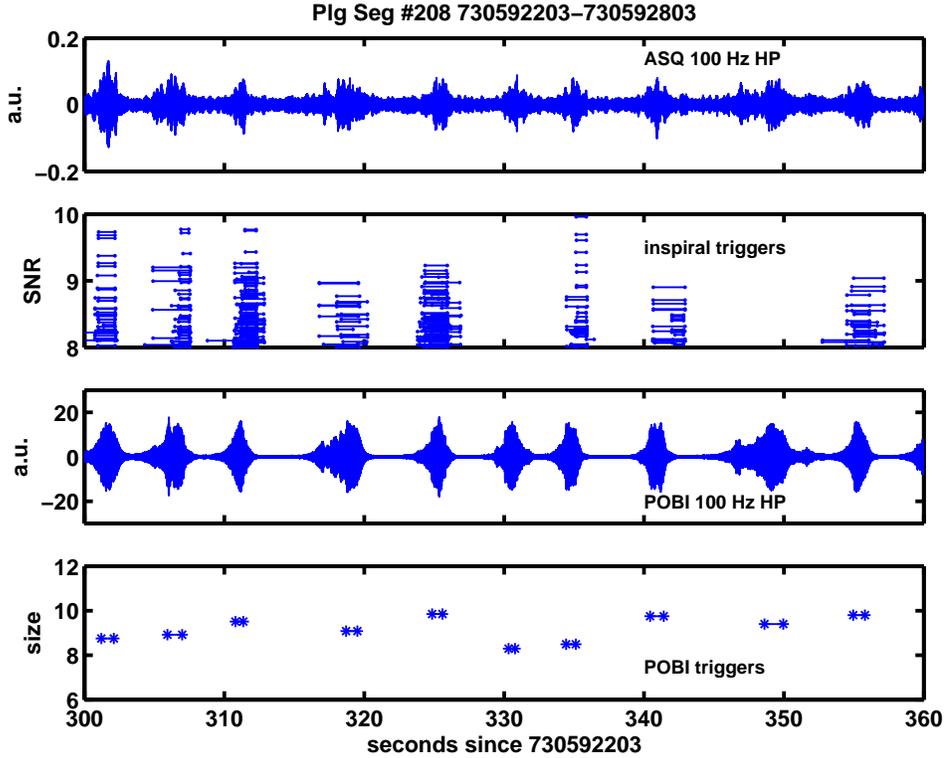}
  \end{center}
\caption{Correlation between glitches in the gravitational wave
channel L1:LSC-AS\_Q (abbreviated ``ASQ'' in the figure) and the
prospective veto channel L1:LSC-POB\_I (``POBI'').  The first and
third plots show time series of these channels after filtering with a
fourth order Chebyshev 100~Hz high-pass filter.  The second and fourth
plots show the time intervals of the triggers reported by the
software, represented as horizontal bars.  In the case of
L1:LSC-AS\_Q, the data was filtered using many template waveforms, and
the SNR for various templates is indicated by the vertical positions
of the bars.  In the case of L1:LSC-POB\_I, the vertical position of
the bar indicates the glitch ``size'' reported by {\it glitchMon}.
The data shown here is from a time in the S2 playground data for which
L1:LSC-AS\_Q is especially glitchy and the efficiency of the
L1:LSC-POB\_I veto is especially good, and is not typical of the
entire S2 run.  } \label{fig4}
\end{figure}

Figure~\ref{fig4} demonstrates the appropriateness of this veto
channel in a different way, using data from an epoch in the S2 run
during which the L1 detector noise was extremely non-stationary.
Presented is a sample time-trace (from the S2 playground data) of the 
interferometer's gravitational wave signal channel, L1:LSC-AS\_Q,
after high-pass filtering, 
along with the signal from L1:LSC-POB\_I. Also displayed in Figure~\ref{fig4}
are the template waveform starting/ending times and the SNR for
the BNS inspiral triggers, and the time intervals of the L1:LSC-POB\_I
veto triggers as reported by {\it glitchMon}.

\begin{figure}[tb]
  \begin{center}
    \includegraphics[width=5.0in,angle=0]{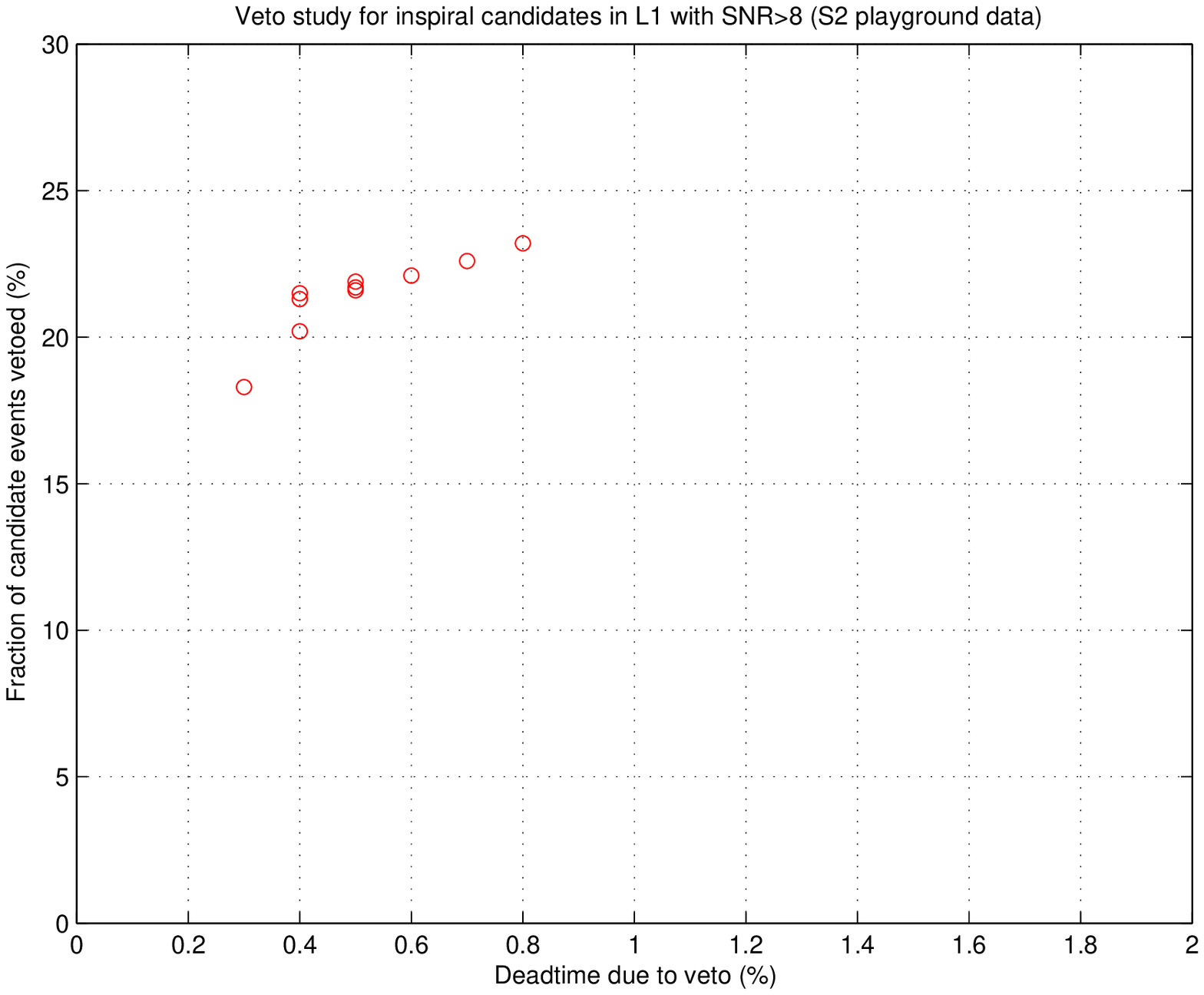}
  \end{center}
\caption{An example of the 
veto efficiency (for BBH inspiral triggers in L1)
versus dead-time for the channel L1:LSC-AS\_Q, 
using symmetric windows of $0.0, \pm 0.05, \pm 0.1, \pm 0.15, 
\pm 0.2, \pm 0.25, \pm 0.3, \pm 0.4, \pm 0.5, \pm 0.75$, and $\pm1.0$~s.
The data from L1:LSC-MICH\_CTRL was filtered with a fourth order Chebyshev $100$~Hz high-pass filter,
and resulting transients with amplitudes exceeding $16 \sigma$ were declared veto triggers.
These results are from the S2 playground data.
} \label{fig5}
\end{figure}

\begin{figure}[tb]
  \begin{center}
    \includegraphics[width=5.0in,angle=0]{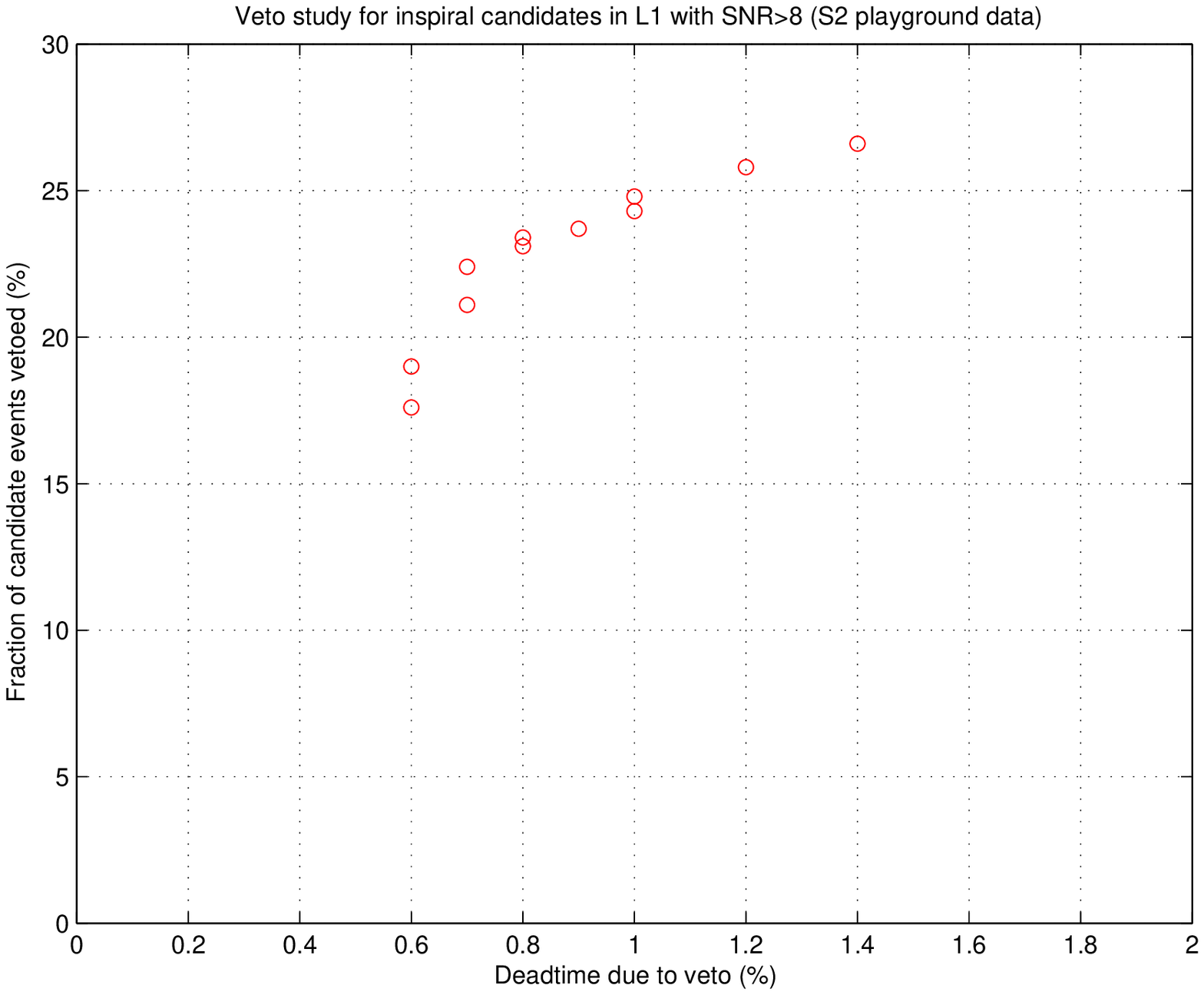}
  \end{center}
\caption{An example of the 
veto efficiency (for BBH inspiral triggers in L1)
versus dead-time for the channel L1:LSC-AS\_Q, 
using symmetric windows of $0.0, \pm 0.05, \pm 0.1, \pm 0.15, 
\pm 0.2, \pm 0.25, \pm 0.3, \pm 0.4, \pm 0.5, \pm 0.75$, and $\pm1.0$~s.
The data from L1:LSC-POB\_I was filtered with a fourth order Chebyshev $70$~Hz high-pass filter,
and resulting transients with amplitudes exceeding $7 \sigma$ were declared veto triggers.
These results are from the S2 playground data.
} \label{fig6}
\end{figure}

In addition to the S2 search for binary neutron star inspiral signals,
a search is underway for binary black hole (BBH) signals.  These
signals have shorter duration and are restricted to a lower frequency
range than in the BNS case, so it is possible that different channels
could provide the best veto conditions.  We have repeated the veto
study using a preliminary list of BBH inspiral triggers in the S2
playground data.  L1:LSC-POB\_I again appears as a good candidate
veto, with efficiency roughly comparable to what was measured for
the BNS case. However, the channel L1:LSC-MICH\_CTRL (the control signal for
the servo loop which controls the differential distance between the
beamsplitter and the input mirrors of the long Fabry-Perot arm
cavitites) appears to yield comparable veto efficiency with
slightly less dead-time.
Figures~\ref{fig5} and~\ref{fig6} show the veto efficiency versus
dead-time for L1:LSC-MICH\_CTRL and L1:LSC-POB\_I, respectively, using
veto time windows up to $\pm 1$~s.
Combining the two channels only increases the veto efficiency by 1~\%,
indicating that the two channels
appear to be glitching concurrently.  The final choice of veto
condition for the BBH inspiral search will be made after refinement of
the inspiral search algorithm and parameters.

\section{Discussion and Conclusions}
\label{disc}
LIGO is now acquiring data, and astrophysically interesting analyses are being 
conducted \cite{LIGO-CW,LIGO-IN,LIGO-BU,LIGO-ST}. From the S1 and S2 data it has been seen
that spurious events, or glitches, can exceed the SNR threshold and occasionally
pass the $\chi^2$ test in the BNS inspiral search. As the interferometers' sensitivities
continue to improve the character of the data changes. The investigations into possible 
vetoes for the inspiral analyses will continue to evolve as the interferometers' performance
changes.

For the S2 inspiral trigger studies we have eliminated problematic data
using data quality checks and a coincident glitch veto.
Data quality cuts eliminate high-noise data in H1 as well as photodiode saturations in all three
LIGO interferometers. Based on preliminary investigations, the low-frequency cutoff for the BNS inspiral search was elevated 
to $100$~Hz in order to avoid problematic non-stationary noise around $70$~Hz. 
The L1:LSC-POB\_I channel provided a moderately efficient veto for the
L1 interferometer, with a dead-time of 3~\%.  No suitable veto
conditions were identified for the H1 or H2 interferometers.

The BBH inspiral search is still being developed and tuned.
Based on preliminary studies, either L1:LSC-POB\_I or
L1:LSC-MICH\_CTRL appears to provide a useful veto, comparable in
efficiency to the BNS case.

For future LIGO science runs we hope to gain a better understanding of
the root causes of glitches. As the 
interferometers' noise decreases it is hoped that environmental causes of triggers will be 
clearly identified. It is likely that low-frequency environmental noise can cause higher frequency
noise in the interferometer output through non-linear coupling. We intend to use higher-order
statistical measures, such as the
{\it bicoherence}, as a means of monitoring the non-linear up-conversion. Also, we hope
to implement further inspiral waveform consistency tests \cite{Shawhan} in order to 
eliminate false triggers that manage to pass the SNR threshold and current $\chi^2$ test.

\verb''\ack
Thanks to Laura Cadonati for providing {\it glitchMon} veto trigger files, 
and to other members of the LSC Burst Analysis Group for discussions. We are
pleased to acknowledge Peter Saulson for carefully reading the manuscript and
providing helpful comments.
This work was supported by grants from the National Science Foundation, 
including grants PHY-0071327, PHY-0107417, PHY-0135389, and PHY-0244357.

\end{document}